\documentclass[12pt]{iopart}

\usepackage[colorlinks=true,linkcolor=red]{hyperref}

\usepackage{amssymb,graphicx,dcolumn,bm,bbm,url,color,setspace}
\usepackage{epstopdf}
\usepackage{fancyhdr}
\usepackage{epsfig}

\expandafter\let\csname equation*\endcsname\relax
\expandafter\let\csname endequation*\endcsname\relax
\usepackage{amsmath}
\usepackage{rotating}
\usepackage{braket}

\newcommand{\je}[1]{#1}

\newcommand{\be}{\begin{equation}}
\newcommand{\ee}{\end{equation}}
\newcommand{\ben}{\begin{eqnarray}}
\newcommand{\een}{\end{eqnarray}}
\newcommand{\bes}{\begin{subequations}}
\newcommand{\ees}{\end{subequations}}
\newcommand{\bF}{\begin{figure}}
\newcommand{\eF}{\end{figure}}

\usepackage{times}

\newcommand{\elem}[2]{\mbox{$|#1\rangle \!\langle #2 |$}}

\definecolor{hcrone}{rgb}{0,1,0}
\definecolor{hcrtwo}{rgb}{1,0,0}
\definecolor{hcrthree}{rgb}{0,0,1}
\definecolor{hcrfour}{rgb}{1,.5,0}

\begin{document}

\title{Recursive quantum detector tomography}

\author{Lijian Zhang$^{1,2}$, Animesh Datta$^1$, Hendrik B. Coldenstrodt-Ronge$^1$, Xian-Min Jin$^{1,3}$, Jens Eisert$^4$, Martin B. Plenio$^{5,6,7}$, Ian A. Walmsley$^1$}

\address{$^1$ Clarendon Laboratory, Department of Physics, University of Oxford, OX1 3PU, United Kingdom}
\address{$^2$ Max Planck Research Department for Structural Dynamics, University of Hamburg-CFEL, Luruper Chaussee 149, 22761 Hamburg, Germany}
\address{$^3$ Center for Quantum Technologies, National University of Singapore, 117543 Singapore}
\address{$^4$ Dahlem Center for Complex Quantum Systems, Freie Universit\"{a}t Berlin, 14195 Berlin, Germany}
\address{$^5$ Institut f\"{u}r Theoretische Physik, Albert-Einstein-Allee 11, Universit\"{a}t Ulm, 89069 Ulm, Germany}
\address{$^6$ Center for Integrated Quantum Science and Technology, Albert-Einstein Allee 11, Universit\"{a}t Ulm, 89069 Ulm, Germany}
\address{$^7$ QOLS, Blackett Laboratory, Imperial College London, London SW7 2BW, UK}
\ead{l.zhang1@physics.ox.ac.uk}

\begin{abstract}

Conventional tomographic techniques are becoming increasingly infeasible for reconstructing the operators of quantum devices of growing sophistication. We describe a novel tomographic procedure using coherent states which begins by reconstructing the diagonals of the operator, and then each successive off-diagonal in a recursive manner. Each recursion is considerably more efficient than reconstructing the operator in its entirety, and each successive recursion involves fewer parameters. We apply our technique to reconstruct the positive-operator-valued measure (POVM) corresponding to a recently developed coherent optical detector with phase sensitivity and number resolution. We discuss the effect of various parameters on the reconstruction accuracy. The results show the efficiency of the method and its robustness to experimental noise.
\end{abstract}

\maketitle

\section{Introduction}
\label{sec:intro}

Quantum detectors inform our classical world of the underlying quantum world through a set of operators known as positive-operator-valued measure (POVM). In practice, the success of many quantum applications rely on certain knowledge of measurement
apparatuses~\cite{Resch07,Datta_ZT-PDSW11,Nick11}. Successful applications of sophisticated detectors rely on a complete and accurate knowledge of the detector, \textit{i.e.} detector characterization. Detector characterization can be implemented in two different ways. One is synthetic, wherein each constituent of a detector is carefully calibrated before being incorporated into a sophisticated physical model of the measurement process. As quantum technologies evolve into increasingly complicated systems, so do quantum detectors, which makes synthetic characterization progressively less feasible. Additionally, any coupling with external degrees of freedom not incorporated into the theoretical model may make the characterization fail~\cite{Fabre11,Zhang12}. A fundamentally different approach is taken in quantum detector tomography (QDT)~\cite{Luis99,ML01,Ariano04,lundeen2008tomography}, where the unknown specifics of a detector are characterized in a largely assumption-free way: here, the POVM of a detector are reconstructed from the outcome statistics in response to a set of tomographically complete certified input states.

To date QDT has been successfully applied to avalanche photodiode (APD)~\cite{Fabre11}, time-multiplexed photon-number-resolving detector (TMD)~\cite{lundeen2008tomography,feito2009measuring,coldenstrodt2009proposed}, transition edge sensors~\cite{Brida11} and superconducting nanowire detectors~\cite{nanotomo}. These detectors are phase-insensitive, \textit{i.e.} they can only measure the mixture of the photon-number states, not the coherence among them. Accordingly, the POVM of these detectors have only non-zero matrix elements on the main diagonals, and the number of parameters to be decided is proportional to $d$. Here $d$ is the dimensionality of the Hilbert space, and for optical detectors can be estimated as the number of photons to saturate the detector. Yet a large number of effects characteristic for quantum mechanics, including entanglement, violation of local realism~\cite{Kuzmich00}, measuring non-classical correlations of radiation fields~\cite{SchukinVogel06}, test of macroscopic realism~\cite{Ourjoumtsev07} \textit{etc.}, relies on quantum coherence. The effort to harness and exploit quantum coherence brings the prosperity of quantum information processing and quantum metrology. Moreover, exploration and utilization of the full Hilbert space of a quantum system requires a detector capable of implementing a tomographically complete set of measurements~\cite{Nunn10}. Therefore optical detectors that can access quantum coherence among photon-number states, \textit{i.e.} phase-sensitive detectors, for example strong- and weak-field homodyne detectors~\cite{Puentes09}, are crucial not only for quantum applications, but also for test of fundamental theories of quantum mechanics. Phase-sensitivity comes with the non-zero off-diagonal matrix elements of the POVM. Thus tomography of a phase-sensitive detector requires the estimation of number of parameters proportional to $d^2$. For practical detectors, $d$ can range from $10^2$ to $10^5$, and $d^2$ from $10^4$ to $10^{10}$. For example, a weak-field homodyne TMD with 9 time bins requires $1.8\times 10^6$ parameters to completely describe its POVM~\cite{Zhang12}, which is about two orders of higher than the largest tomography that had been performed until then~\cite{8qubit_05}. Such large set of parameters represents a considerable challenge to the characterization of phase-sensitive detectors.

In this work we explore potential solutions to QDT of phase-sensitive quantum detectors. In particular we introduce an algorithm that allows to reconstruct the POVM recursively, with no more than $d$ parameters per recursion. Simulations with the QDT of weak-field detectors demonstrate the robustness of this algorithm.

\section{Definition of the problem}
\label{sec:def}
QDT is performed by preparing a set of known probe states $\{\rho_m\}$ incident on a quantum detector and observing the detector outcomes. The probability of registering outcome $n$ is given by the Born rule
\begin{equation}
p_{n|m} = \tr(\rho_m \Pi_{n}),
\label{eq:Born}
\end{equation}
where $\{\Pi_{n}\}$ is the POVM of the detector with $n=0,\dots  ,N-1$, and $N$ is the number of possible outcomes of the detector. In practice the experiment is repeated for each of many identical copies of the probe states, and the frequency $f_{n|m}$ for each measurement outcome $n$ occurring when probe state $\rho_m$ is used is recorded. Then $p_{n|m}$ can be estimated from the relative frequency $p_{n|m} = f_{n|m}/\sum_{n} f_{n|m}$. One can then invert Eq.~(\ref{eq:Born}) to find $\Pi_n$. For a finite number of repetitions, there are always fluctuations in the estimation of $p_{n|m}$, therefore the inversion should normally be preformed with convex optimization.

A key requirement is that the set of probe states must be \emph{tomographically complete}. However, it is also important that the set of probe states are experimentally feasible. That means that the states should themselves be well-characterized, and that a large variety should be available with high precision. There are proposed methods to generate the probe states through quantum correlations~\cite{Ariano04, Brida12}. Yet with current quantum optical sources it is very hard to generate the probe states strong enough to saturate the detector under test. For photodetector measurements, there is a more straightforward option. The set of coherent state vectors $| \alpha \rangle$ of an optical beam are ideal candidates, where $\alpha = |\alpha|e^{i\theta}$ is the complex amplitude. They are overcomplete in the sense that two different coherent states are not orthogonal with each other yet any quantum state can be decomposed on the set of coherent states. Therefore coherent states can form a tomographically complete set by transforming their amplitude (by means of optical attenuation) and their phase (with a simple delay line). Importantly, they are generated very easily by a laser.

With coherent states as input, the probabilities are given by
\begin{equation}
p_{n|\alpha}= \langle \alpha | \Pi_n | \alpha \rangle = \pi Q_n (\alpha),
\label{eq:Qfn}
\end{equation}
where $Q_n(.)$ is the $Q$-function of the detector POVM elements $\Pi_n$. This is equal to the Husimi representation of the POVM, and is uniquely and invertibly related to the POVM. To reconstruct $\Pi_n$ one can write both $|\alpha\rangle$ and $\Pi_n$ in the photon-number basis and truncate the expansion at $d-1$, where $d-1$ is the number of photons that saturate the detector.
\begin{eqnarray}
|\alpha\rangle & = & e^{-|\alpha|^2/2}\sum_{j=0}^{d-1} \frac{|\alpha|^j}{\sqrt{j}}e^{i j\theta} |j\rangle, \label{eq:coh_expand}\\
\Pi_n & = & \sum_{j,k=0}^{d-1} \pi_{n}^{j,k} |j\rangle\langle k|. \label{eq:Pi_expand}
\end{eqnarray}
Then Eq.~(\ref{eq:Qfn}) can be written as
\begin{equation}
p_{n|\alpha} = e^{-|\alpha|^2}\sum_{j,k=0}^{d-1} \frac{|\alpha|^{j+k}}{\sqrt{j!k!}}e^{i\left(k-j\right)\theta}\pi_{n}^{j,k}.
\label{eq:cp_expand}
\end{equation}
We can relabel Eq.~(\ref{eq:cp_expand}) in $s = kd+j+1$ ($1\leq s \leq d^2$), with $j = [(s-1) \textrm{ mod } d]$ and $k = (s-j-1)/d$. For $M$ probe states, there are $M\times N$ linear equations, which can be written in a matrix form
\begin{equation}
P = F\tilde{\Pi}
\label{eq:bohn_mat}
\end{equation}
where $P$ is an $M\times N$ matrix with elements $P^{m,n} = p_{n|\alpha_m}$, $F$ is an $M\times d^2$ matrix with elements
\begin{equation}
F^{m,s} = e^{-|\alpha|^2} \frac{|\alpha_m|^{j(s)+k(s)}}{\sqrt{j(s)!k(s)!}}e^{i\left(k(s)-j(s)\right)\theta_m},
\label{eq:f_ele}
\end{equation}
and $\tilde{\Pi}$ is a $d^2\times N$ matrix with elements $\tilde{\pi}^{s,n} = \pi_{n}^{j(s),k(s)}$. In practice where the experimental noise is taken into account, the POVM set can be estimated from Eq.~(\ref{eq:bohn_mat}) with convex optimization subject to the constraints
\begin{eqnarray}
\Pi_n \geq  0, \label{eq:cons_pos} \\
\sum_{n=0}^{N-1} \Pi_n  =  I, \label{eq:cons_ind}
\end{eqnarray}
where $I$ is the identity operator. One common approach is the least square estimation
\begin{equation}
\min ||P-F\tilde{\Pi}||_2,
\label{eq:opt_wo_reg}
\end{equation}
where $||A||_2 = \sqrt{\textrm{Tr}(A^\dag A)}$ is the Frobenius norm. The reconstruction problem effectively deconvolves a coherent state from the statistics to obtain the POVM set. This is an ill-conditioned problem, as seen by the large ratio between the largest and smallest singular values of the matrix $F$. This makes the POVM extremely vulnerable to small fluctuations in the measurement statistics. Such instability can be overcome by adding a regularization function $g(\tilde{\Pi})$ to the optimization~\cite{lundeen2008tomography,feito2009measuring}, therefore the problem is modified as
\begin{eqnarray}
\min \{||P-F\tilde{\Pi}||_2 + g(\tilde{\Pi})\}, \nonumber \\
\mbox{subject to~~~} \Pi_n \geq 0, && \sum_{n=0}^{N-1}\Pi_n = I.
\label{eq:sdp}
\end{eqnarray}
For a phase-insensitive detector with finite detection efficiency, one would expect the variation of the diagonal matrix elements to be smooth, therefore a regularization function known as Tikhonov regularization~\cite{BoydVandenberge04} is applied
\begin{equation}
\label{eq:regcond}
g(\tilde{\Pi})=\gamma \sum_{j,n}|\pi_n^{j,j}-\pi_n^{j+1,j+1}|^2.
\end{equation}
This limits the variation between adjacent elements along the diagonal matrix elements. Yet for a phase-sensitive detector a regular function is not easy to find: even as each of the leading diagonals are smooth, the relation among different leading diagonals can be arbitrary.

An alternative approach for convex optimization is maximum likelihood estimation, which was also proposed for QDT~\cite{ML01}. Maximum likelihood alleviates the requirement of the regularization function. However, its convergence speed is normally not high. Moreover, both the maximum likelihood estimation and the least square estimation in Eq.~(\ref{eq:sdp}) requires the reconstruction of the whole POVM matrices at the same time. When the size of the matrices becomes large, the problem becomes infeasible. For example, the estimation of a POVM set with 9 elements each of which is a 50 by 50 matrix is already a hard problem for the capability of current multi-processor desktops (2xQuad Core 3GHz, 8GB RAM).

The engineering of large entangled quantum states and development of sophisticated quantum operations has set a challenge for standard quantum tomography techniques. There has been increased interest in the development of novel algorithm with improved efficiency for special situations. In particular, there are process tomography schemes that allows to selectively reconstruct the state or process matrix partially in each run. Several of them use a-priori knowledge about the state such as their symmetry~\cite{Toth10,Cramer10,Baumgratz12}, or simply reconstructing a subset of the entire state or process~\cite{Bendersky08,Schmiegelow_CLP10}. Using improved techniques from classical signal processing have also become common, such as compressed sensing~\cite{Gross10,Shabani11}. 
In the following, we introduce a novel algorithm that reconstructs the POVM elements recursively in multiple runs.

\section{The detector model}
\label{sec:det_mod}
\begin{figure}[ht]
\begin{center}
\includegraphics[width=0.95\textwidth]{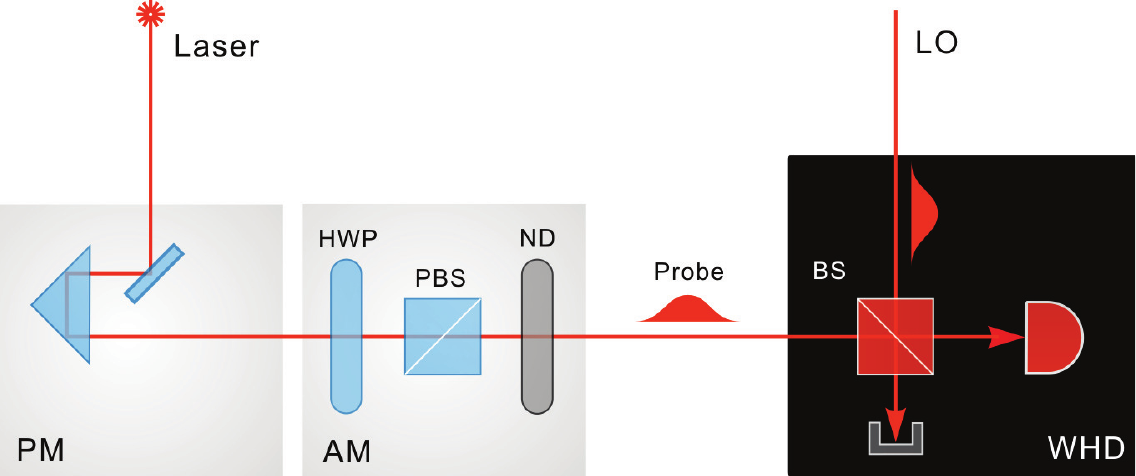}
\caption[]{The configuration of a weak-field homodyne detector and its tomography setup. A set of probe states are prepared by the phase-modulation (PM) and amplitude-modulation (AM) of the output of a laser. The magnitude of the probe state is adjusted by a half-wave plate (HWP) followed by a polarizing beam-splitter (PBS) and neutral density filters (ND). The phase of the probe state is controlled by a piezo translator. The setup of the weak-field homodyne detector (WHD) is shown in the black box.}
\label{fig:setup}
\end{center}
\end{figure}

The algorithm discussed in this work can be universally applied to the tomography of any quantum detector. To illustrate its working in this work we consider a simple example of phase-sensitive detector, the weak-field homodyne APD. The configuration of such detectors together with a schematic tomography setup is given in Fig.~(\ref{fig:setup}). The probe state is prepared by the phase modulation and amplitude modulation of the output of a laser system. The weak-field homodyne detector is shown in the black box, where the input state interferes with a local oscillator (LO) and detected by a photon-number-resolving or photon-counting detector. For a weak-field homodyne APD, there are two possible measurement outcomes, no-click and click events, and the corresponding POVM elements $\Pi_0$ and $\Pi_1$ are given as~\cite{Puentes09}.
\begin{eqnarray}
\Pi_0 & = & \sum_{c=0}\sum_{d=0} (1-\eta_{\rm APD})^c \frac{e^{-|\alpha_{L}|^2}}{c!d!2^{c+d}}  \nonumber \\
& & \times(\alpha_L^{\ast} + \hat{a}^{\dag})^c (\alpha_L^{\ast}-\hat{a}^{\dag})^d |0\rangle \langle 0| (\alpha_L + \hat{a})^c (\alpha_L-\hat{a})^d, \label{eq:povm_0} \\
\Pi_1 & = & I - \Pi_0. \label{eq:povm_1}
\end{eqnarray}
where $\alpha_L$ is the complex amplitude of the LO and $\eta_{\rm APD}$ is the detection efficiency of APD.

\section{A selective algorithm with Glauber-Sudarshan $P$-function}
\label{sec:pfun}
Before we proceed to the recursive algorithm, we consider a more straightforward selective algorithm. Each matrix element $\pi_{n}^{j,k}$ of $\Pi_{n}$ is given by
\begin{equation}
\pi_{n}^{j,k} = \textrm{Tr}\left( |k\rangle\langle j| \Pi_n\right).
\label{eq:mat_ele}
\end{equation}
Using the Glauber-Sudarshan decomposition of $|k\rangle\langle j|$
\begin{equation}
|k\rangle\langle j|  = 2\int P^{j,k}(\alpha) |\alpha\rangle\langle \alpha| d^2 \alpha,
\label{eq:pfun}
\end{equation}
we have
\begin{equation}
\pi_{n}^{j,k} = 2\int P^{j,k}(\alpha) \langle\alpha | \Pi_n |\alpha\rangle d^2 \alpha = 2\pi \int P^{j,k}(\alpha) Q_n(\alpha) d^2 \alpha.
\label{eq:pqfun}
\end{equation}
In principle we can estimate each individual matrix element $\pi_{n}^{j,k}$ separately with either the exact form of $P^{j,k}(.)$ which contains the derivative of Dirac-delta function or the approximated regular form~\cite{Klauder66}. Similar method has been used for quantum process tomography~\cite{Lobino2008, Lobino09, rahimi2011quantum}. However, due to the singularity of $P$-function, this scheme is extremely sensitive to the noise in the measured $Q$-function of the POVM element, rendering it infeasible for practical QDT.

As an example, we consider the reconstruction of no-click POVM of a weak-homodyne APD with the reflectivity of the beam-splitter of 0.5, LO intensity $|\alpha_{LO}|^2 = 5$ and quantum efficiency of the APD $60\%$ (overall detection efficiency $30\%$). We choose the probe states that sample the phase space from $X,Y = -10$ to $X,Y = 10$ with a step size 0.05, where $X$ and $Y$ are the two quadratures of an optical field. We assume that for each probe state we run the experiment $f$ times, then the expected frequency to get the no-click event is $\langle f_{0|\alpha}\rangle = \pi Q_{0}(\alpha) f$. In practice there are many experimental imperfections that may induce fluctuations to the measurement results. In this work we only consider the most fundamental fluctuation due to the random nature of the outcome of each measurement process, and simulate it by assuming that $f_{0|\alpha}$ is a random number with a binomial distribution, and assigning the experimentally measured $Q$-function as $Q^{exp}(\alpha) = f_{0|\alpha}/f$.
\begin{figure}[ht]
\begin{center}
\includegraphics[width=0.9\textwidth]{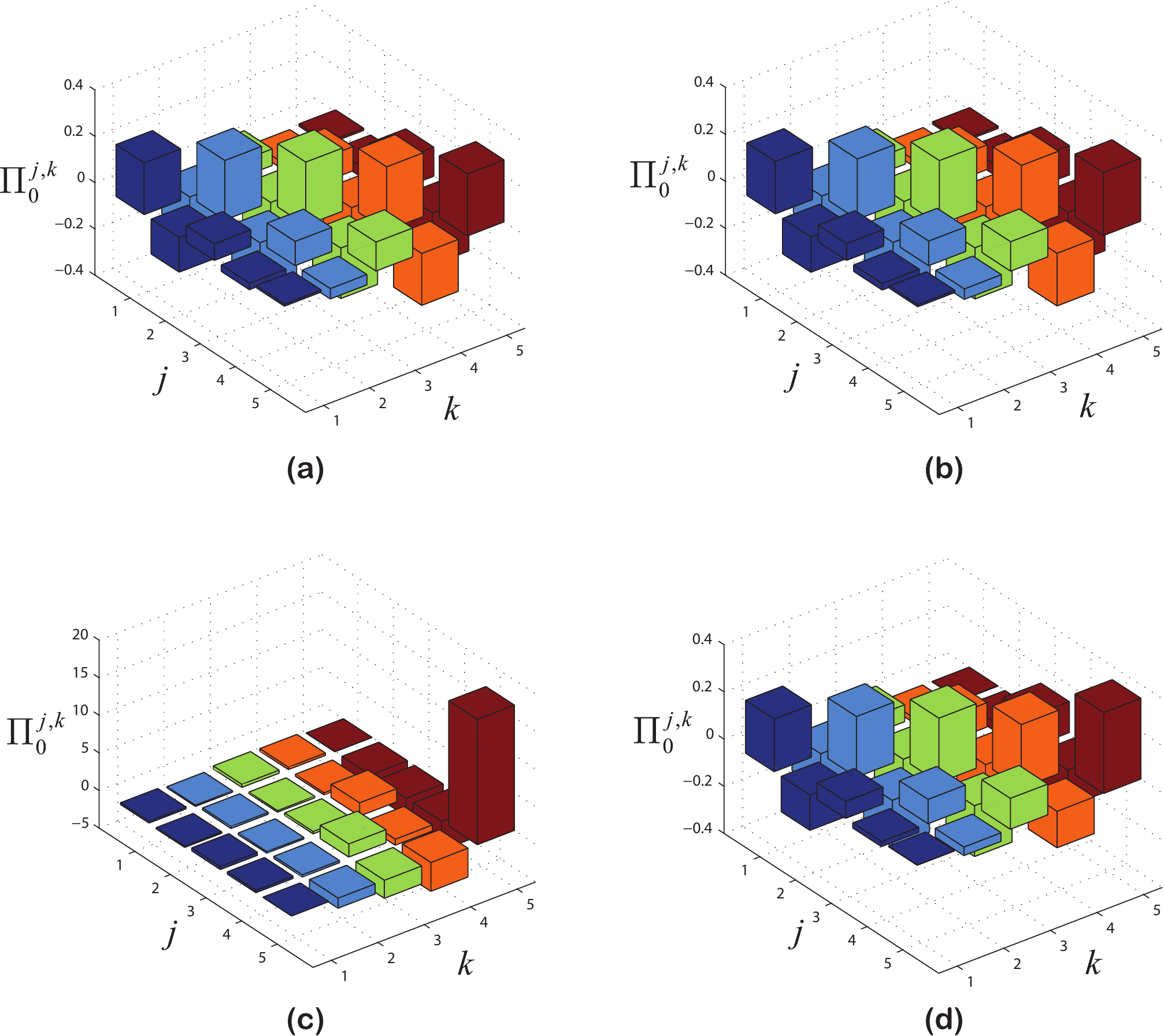}
\caption[]{Simulation results for the tomography of the no-click POVM of a weak-field homodyne APD, displayed up to $d=5$. From (a) to (d): theoretical POVM; reconstructed without noise; $f=10^5$; $f=10^{10}$.}
\label{fig:tomo_p_fun}
\end{center}
\end{figure}

The results are shown in Fig.~(\ref{fig:tomo_p_fun}). Without experimental fluctuations, the reconstructed POVM matches almost perfectly with the theoretical prediction. Yet when there is the presence of experimental noise, the results deviates from the theoretical prediction very quickly: for $f=10^5$ as in Fig.~(\ref{fig:tomo_p_fun}(c)), the reconstructed POVM element is not even a physical measurement operator. Only when the number of measurement is large enough $f=10^{10}$ and thus the experimental fluctuations is small, the reconstructed POVM element is close to the real one. The results presented here is reconstructed up to the 4-photon component. The P-functions of higher photon-components are more singular, and the reconstruction is more sensitive to experimental fluctuations. Therefore, this method has a serious problem for its scalability. For practical QDT, we need to seek another solution.

\section{Recursive reconstruction of the POVM set}
\label{sec:recur}

\subsection{Outline of the recursive reconstruction}
In this section, we discuss a novel recursive method for the tomographic reconstruction of quantum operators. We begin with Eq.~(\ref{eq:cp_expand}). Before relabeling, we first integrate over the probe state phase $\theta$. With
\begin{equation}
\int_{0}^{2\pi} e^{i(k-j)\theta} d\theta = 2\pi \delta_{k,j},
\label{eq:int_theta}
\end{equation}
we have
\begin{equation}
\frac{1}{2\pi}\int_{0}^{2\pi} p_{n|\alpha} d\theta = \sum_{j=0}^{d-1} e^{-|\alpha|^2} \frac{|\alpha|^{2j}}{j!} \pi_{n}^{j,j}.
\label{eq:diag}
\end{equation}
The left side of Eq.~(\ref{eq:diag}) is a partial integration of the experimental results, while the right side involves only the main diagonals of the POVM. Eq.~(\ref{eq:diag}) can be interpreted as using phase-randomized coherent states as input to the detector. Since the input states are completely mixed, the measurement process only involve the main diagonals of the POVM. In a practical experiment, one should change the integration on the left side to summation. The probe states should be prepared with $M_{a}$ different amplitudes and for each amplitude there are $M_{p}$ different phases. In total there are $M = M_{a} M_{p}$ probe states with the complex amplitudes $\alpha_{u,v} = |\alpha_u| e^{i\theta_{u,v}}$, with $u = 1 ,\dots  ,N_a$ and $v = 1 ,\dots  ,N_p$. Therefore the integration on the left side of Eq.~(\ref{eq:diag}) can be approximated as
\begin{equation}
\frac{1}{2\pi}\int_{0}^{2\pi} p_{n|\alpha} d\theta \approx \frac{1}{M_p} \sum_{v = 1}^{M_p} p_{n|\alpha_{u,v}}.
\label{eq:int_sum}
\end{equation}
For reconstructing the off-diagonals we first note that POVM elements are Hermitian and it is sufficient to reconstruct just the upper or lower off-diagonals. We multiply Eq.~(\ref{eq:cp_expand}) by $e^{-il\theta}$ and integrate over $\theta$. Since
\begin{equation}
\int_{0}^{2\pi} e^{i(k-j)\theta} d\theta = 2\pi \delta_{k,j},
\label{eq:int_theta_2}
\end{equation}
we have
\begin{equation}
\frac{1}{2\pi}\int_{0}^{2\pi} p_{n|\alpha}e^{-il\theta} d\theta = \sum_{j=0}^{d-1} e^{-|\alpha|^2} \frac{|\alpha|^{2j+l}}{\sqrt{j!(j+l)!}} \pi_{n}^{j,j+l}.
\label{eq:offdiag}
\end{equation}
Again for practical experiment the integration should be substituted with a summation
\begin{equation}
\frac{1}{2\pi}\int_{0}^{2\pi} p_{n|\alpha}e^{-il\theta} d\theta \approx \frac{1}{M_p} \sum_{v = 1}^{M_p} p_{n|\alpha_{u,v}}e^{-il\theta_{u,v}}.
\label{eq:int_sum_2}
\end{equation}
with an error
\begin{equation}
\Delta = -\frac{2 \pi^3}{3 M_p^2} \frac{d^2 \left(p_{n|\alpha}e^{-il\theta}\right)}{d\theta^2} \sim -\frac{2 \pi^3 l^2}{3 M_p^2}.
\label{eq:nume_err}
\end{equation}
Eq.~(\ref{eq:offdiag}) includes the situation in Eq.~(\ref{eq:diag}) when $l=0$. For each $l$, there are $M_a$ equations. As has been done in Eq.~(\ref{eq:bohn_mat}), we can write them in a matrix form $P^{(l)}=F^{(l)}\tilde{\Pi}^{(l)}$, with $P^{(l)}$ an $M_a\times N$ matrix, $F^{(l)}$ an $M_a\times d$ matrix, $\tilde{\Pi}^{(l)}$ a $d \times N$ matrix, and the coefficients given by Eq.~(\ref{eq:offdiag}). Comparing with Eq.~(\ref{eq:bohn_mat}), all the matrices involved here are significantly smaller. With the presence of the experimental fluctuations, the reconstruction becomes a convex optimization problem, in fact a semi-definite problem,
\begin{eqnarray}
\min \{||P^{(l)}-F^{(l)}\tilde{\Pi}^{(l)}||_2 &+& g(\tilde{\Pi}^{(l)})\}, \nonumber \\
\mbox{subject to~~~} \Pi_n \geq 0, && \sum_{n=0}^{N-1}\Pi_n = I.
\label{eq:sdp_rec}
\end{eqnarray}
Since this is a convex optimization problem, there is only one minimum value which can be found with YALMIP toolbox for Matlab~\cite{YALMIP} with the solver SeDuMi~\cite{sedumi} utilizing primal-dual interior point methods~\cite{BoydVandenberge04}. For the examples discussed in this paper, the calculation converges to its minimum in less than 40 iterations which takes about 1 seconds on a multi-process desktop (Dual Core 2GHz, 2GB RAM). This allows us to reconstruct the POVM recursively for $l=0,\dots  ,d$. For $l=0$, the second condition is that the summation of the main diagonals of all the POVM elements equals to 1, while for $l\neq 0$, this condition means that the summation of the $l$th leading diagonals of all the POVM elements equals 0. The positivity condition should be enforced recursively based on Sylvester's criterion, which states that a matrix is positive if and only if all of its principal minors are positive. For $l=0$ this requires all the matrix elements on the main diagonals to be positive. Now we derive the condition for $l>0$. We start with $l=1$. In Eq.~(\ref{eq:cons_step_1}) we show the matrix $\Pi_{n}$ where the diagonal elements (green) have been determined using Eq.~(\ref{eq:diag}) and the first row of off-diagonals (red) is to be determined with Eq.~(\ref{eq:int_sum_2}). Any other entry is unknown and will not be reconstructed at this step.
\begin{spacing}{1}
\begin{equation}
\Pi_{n}=
\begin{pmatrix}
{\color{hcrone}\pi_{n}^{0,0}} & \negmedspace{\color{hcrthree}\text{---}} & \negmedspace{\color{hcrtwo}\pi_{n}^{0,1}} & & \text{?} & & \text{?} & \hdots & &   \\
{\color{hcrthree}|} & & {\color{hcrthree}|} & & & & & &  &  \\
{\color{hcrtwo}{\pi_{n}^{0,1}}^\ast} & {\color{hcrthree}\text{---}} & {\color{hcrone}\pi_{n}^{1,1}} & {\color{hcrfour}\text{---}} & {\color{hcrtwo}\pi_{n}^{1,2}} & & \text{?} & &   &\\
& & {\color{hcrfour}|} & & {\color{hcrfour}|} & & & & &   \\
\text{?} & & {\color{hcrtwo}{\pi_{n}^{1,2}}^\ast} & {\color{hcrfour}\text{---}} & {\color{hcrone}\pi_{n}^{2,2}} & & \ddots & & &   \\
& & & & & & & &   &\\
\text{?} & & \text{?} & & \ddots & & \ddots &  & &  \\
\vdots & & & & & &  &  & & \\
& & & & &   & & {\color{hcrone}\pi_{n}^{d-2,d-2}} & \text{---} & {\color{hcrtwo}\pi_{n}^{d-2,d-1}} \\
& & & & &   & & | & & |  \\
& & & & &   & & {\color{hcrtwo}{\pi_{n}^{d-2,d-1}}^\ast} & \text{---} & {\color{hcrone}\pi_{n}^{d-1,d-1}}
\end{pmatrix}.
\label{eq:cons_step_1}
\end{equation}
\end{spacing}
For an input state vector of the form $\ket{\Phi}= a\ket{j}+b \ket{j+1}$ the effective submatrix of $\Pi_{n}$ is given by
\begin{equation}
\Pi^{j,1}_{n}=
\begin{pmatrix}
\pi_{n}^{j,j} & \pi_{n}^{j,j+1} \\
{\pi_{n}^{j,j+1}}^\ast & \pi_{n}^{j+1,j+1}
\end{pmatrix},
\label{equ:minimatrixTWO}
\end{equation}
which needs to be positive. We marked these submatrices with {\color{hcrthree} blue}, {\color{hcrfour} orange}, and black lines for the $j=0$, $j=1$, and $j=M-1$ cases in Eq.~(\ref{eq:cons_step_1}). Thus we imposed the additional constraint that all diagonally centered $2\times2$ submatrices of $\Pi^{j,1}_{n}$ need to be positive for the reconstruction of the first off-diagonal. This condition is satisfied if and only if the determinant $\textrm{det} (\Pi^{j,1}_{n})$ is positive, which implies
\begin{equation}
\left|\pi_{n}^{j,j+1}\right|^2 \leq \pi_{n}^{j,j}\pi_{n}^{j+1,j+1} \,\,\,\,\forall n,j.
\label{eq:pos_cond_1}
\end{equation}

For the following reconstruction of the $l$th leading diagonal we impose a similar constraint on the $\left(1+l\right)\times\left(1+l\right)$ submatrices start with $\pi_{n}^{j,j}$, illustrated in Eq.~(\ref{equ:constraintmore})
\begin{spacing}{1}
\begin{equation}
\Pi^{j,l}_{n}=
\begin{pmatrix}
& &{\color{hcrone}\pi_{n}^{j,j}} & \negmedspace{\color{hcrthree}\text{---}} & \cdots & \negmedspace{\color{hcrthree}\text{---}} & \cdots & \negmedspace{\color{hcrtwo}\pi_{n}^{j,j+l}} & & \\
& & {\color{hcrthree}|} & & & & & {\color{hcrthree}|} & & \\
& & \vdots & & & & & \vdots & &\\
& & {\color{hcrthree}|} & & & & & {\color{hcrthree}|} & &\\
& & {\color{hcrone}{\pi_{n}^{j,j+l}}^\ast} & \negmedspace{\color{hcrthree}\text{---}} & \cdots & \negmedspace{\color{hcrthree}\text{---}} & \cdots & \negmedspace{\color{hcrtwo}\pi_{n}^{j+l,j+l}} & &\\
\end{pmatrix},
\label{equ:constraintmore}
\end{equation}
\end{spacing}
\vspace{.45cm}
where only $\pi_{n}^{j,j+l}$ is unknown. It is required that $\Pi^{j,l}_{n}$ is a positive matrix. Since constraints in the previous steps ensure all its principle minors are positive, this condition is equivalent to that its determinant $\textrm{det}(\Pi^{j,l}_{n})$ is positive. To facilitate numerical calculations we derive the bounds on $\pi_{n}^{j,j+l}$, which can be done by noticing that $\textrm{det}(\Pi^{j,l}_{n})$ is a quadratic polynomial of $\pi_{n}^{j,j+l}$
\begin{equation}
\textrm{det}(\Pi^{j,l}_{n}) = A \times (\pi_{n}^{j,j+l})^2 + B \times \pi_{n}^{j,j+l} + C.
\label{eq:det_poly}
\end{equation}
It is easy to see that $A$ is positive, since $A$ is the product of the elements along the anti-diagonal and $\Pi^{j,l}_{n}$ is Hermitian. Therefore $\textrm{det}(\Pi^{j,l}_{n})\geq 0$ implies that
\begin{equation}
\frac{-B-\sqrt{B^2-4AC}}{2A} \leq \pi_{n}^{j,j+l} \leq \frac{-B+\sqrt{B^2-4AC}}{2A} \textrm{~~~if~~~} B^2-4AC \geq 0.
\label{eq:bound_lth}
\end{equation}
The value of $A$, $B$ and $C$ can be easily estimated from Eq.~(\ref{eq:det_poly}) by substituting $\pi_{n}^{j,j+l} = \pm 1,0$ into $\Pi^{j,l}_{n}$ and calculating the determinant numerically.

\subsection{The number of leading diagonals $l$}
\label{sec:l}
To reconstruct the full POVM matrices, we should run the calculation in Eq.~(\ref{eq:sdp_rec}) until $l = d-1$. As can be seen from Eq.~(\ref{eq:nume_err}), for higher $l$ it requires increased number of phases $M_p$ to reduce the numerical error. Yet, in practice the number of leading diagonals can be estimated during the calculation. From the positivity condition, one has
\begin{equation}
|\pi_{n}^{j,j+l} |^2\leq {\pi_{n}^{j,j} \pi_{n}^{j+l,j+l}}.
\label{eq:n_diag_1}
\end{equation}
Therefore, after the reconstruction of the principle diagonals, we can put a bound on the number of leading diagonals to be reconstructed. Moreover, in any practical detector there is always a finite fluctuation of the reference phase (with a fluctuation of $2\pi$ for a phase-insensitive detector), which will further reduce the number of leading diagonals, as shown below. In fact, this phase noise will ensure
that the entries of the POVM elements decay exponentially away from their main diagonal.

Assume the reference phase has a Gaussian distribution with width of $\delta>0$. Instead of having a POVM $\Pi_n$ we have
\begin{equation}
\Pi'_n = \frac{1}{\delta\sqrt{2\pi}}\int_{-\pi}^{\pi} d\xi R(\xi)^{\dag}\Pi_n R(\xi) \exp(-\xi^2/(2\delta^2)),
\label{eq:povm_phase_fluc}
\end{equation}
where $R(\xi)$ is the rotation operator in phase space with angle $\xi$. The matrix elements of $\Pi'_n$ are given by
\begin{eqnarray}
\label{eq:povm_phase_fluc_2}
\pi_n^{'j,j+l} &=& \frac{1}{\delta\sqrt{2\pi}} \int_{-\pi}^{\pi} d\xi \langle j |R(\xi)^{\dag}\Pi_n R(\xi) |j+l\rangle \exp(-\xi^2/(2\delta^2)) \nonumber \\
&=& \frac{1}{\delta\sqrt{2\pi}} \int_{0}^{2\pi} d\xi \langle j |\Pi_n |j+l\rangle \exp(-\xi^2/(2\delta^2)+il\xi) \nonumber \\
&=&  \frac{\pi_n^{j,j+l}}{\delta\sqrt{2\pi}} \int_{-\pi}^{\pi} d\xi
\exp(-\xi^2/(2\delta^2)+il\xi) \nonumber \\
&=& \frac{\pi_n^{j,j+l}}{\delta\sqrt{2\pi}} \int_{-\pi}^{\pi}
\je{d\xi}
\exp(-\xi^2/(2\delta^2))\cos(l\xi).
\end{eqnarray}
Intuitively,
if the fluctuation of the phase reference is small, \textit{i.e.}, $\delta \ll \pi$, the last integration in Eq.~(\ref{eq:povm_phase_fluc_2}) can be approximated as
\begin{equation}
\label{eq:int_fluc}
\int_{-\pi}^{\pi} \je{d\xi} \exp(-\xi^2/(2\delta^2))\cos(l\xi) \approx \int_{-\infty}^{\infty} \je{d\xi}
\exp(-\xi^2/(2\delta^2))\cos(l\xi) = \delta \sqrt{2\pi} \exp(-l^2 \delta^2/2).
\end{equation}
The intuition of exponentially decaying coefficients can be made rigorous as follows. One has, for w.l.o.g.\ $l$ even,
\begin{eqnarray}
	\left| \int_{-\pi}^{\pi} \je{d\xi} \exp(-\xi^2/(2\delta^2))\cos(l\xi)\right|&\leq&
	\delta \sqrt{2\pi} \exp(-l^2 \delta^2/2)	\nonumber\\
	& & +2
	\left| \int_{\pi}^{\infty} \je{d\xi} \exp(-\xi^2/(2\delta^2))\cos(l\xi)\right|\nonumber\\
	&=&
	\delta \sqrt{2\pi} \exp(-l^2 \delta^2/2)	\nonumber\\
	& & +2
	\left| \int_{0}^{\infty} \je{d\xi} \exp(-(\xi+\pi)^2/(2\delta^2))\cos(l\xi)\right|\nonumber\\
	&\leq&
	\delta \sqrt{2\pi} \exp(-l^2 \delta^2/2)	\nonumber\\
	& & +2
	\left| \int_{0}^{\infty} \je{d\xi} \exp(-\xi^2/(2\delta^2))\cos(l\xi)\right|\nonumber\\
	&=&
	2
	\delta \sqrt{2\pi} \exp(-l^2 \delta^2/2)	.
\end{eqnarray}
Thus, the matrix elements of $\Pi'_n$ satisfy
\begin{equation}
	|\pi_n^{'j,j+l} |\leq 2|\pi_n^{j,j+l}
	|
	  \exp(-l^2 \delta^2/2).
\label{eq:povm_fluc_3}
\end{equation}
The $l$th leading diagonal is decreased by a factor of $2\exp(-l^2 \delta^2/2)$. With the increase of $l$ this factor increases therefore reduces the number of significant leading diagonals in $\Pi'_n$, leading to $l \ll d$. That is to say, the
effort of reconstruction up to a constant error is of order $O(d)$ instead of $O(d^2)$.
For example, with a phase fluctuation of $10$ degrees the 18th leading diagonal is reduced to $1\%$ of that with no LO phase fluctuation.

Another reason for the reduction of the required calculation for the leading diagonals comes from one of the major point of performing detector tomography: to predict the response of the detector with various input quantum states. For situations involving input states with a fixed photon number $N$, like $N00N$ states~\cite{Sanders_89} or Holland-Burnett states~\cite{Datta_ZT-PDSW11}, we only require $N$ leading diagonals of the POVM elements to predict all measurement outcomes. Due to the lack of bright quantum sources, $N$ is usually small (less than 8).

\subsection{Regularization}
\label{sec:reg}

The numerical stability of a reconstruction algorithm is one of its vital certificates. Numerical instability has been a common problem in tomography~\cite{Boulant_HP_Cory03,Jezek_F_Hradil03}, particularly so in using phase space data from homodyne tomography to reconstruct operators in the Fock spaces~\cite{Lvovsky_Raymer09}. Tools such as pattern functions~\cite{Leonhardt95,Dariano95,Wunsche97} exist that can bridge this gap. They are however, hard to identify and cumbersome to work with~\cite{Leonhardt96}. The use of maximum likelihood functions has also been suggested for detector tomography~\cite{ML01,Ariano04}. Unfortunately, as mentioned earlier, the speed of the convergence of such algorithms is not generally guaranteed to be high, becoming exponentially slow for certain problems.

We strike a balance by developing a recursive algorithm that is efficient by virtue of being cast as a semi-definite programme, as is evident from the convex function to be minimized, and the linear constraints in Eq.~(\ref{eq:sdp}). Unfortunately, this still leaves us with an ill-conditioned problem, primarily due to extremely large ratio between the largest and the smallest singular values of the matrix $F^{(l)}$. This is a consequence of the large range of coherent state amplitudes needed to cover the entire dynamical range of the detector in the Fock space. The most common outcome of this ill-conditioning is to result in reconstructed POVMs that have sharp discontinuities~\cite{feito2009measuring}. As shown in Eq.~(\ref{eq:regcond}) this can be resolved by a smoothing function or Tikhonov regularization~\cite{BoydVandenberge04}. We will next discuss how this mathematical technique is physically enforced in realistic detectors.

\begin{figure}[ht]
\begin{center}
\includegraphics[width=0.95\textwidth]{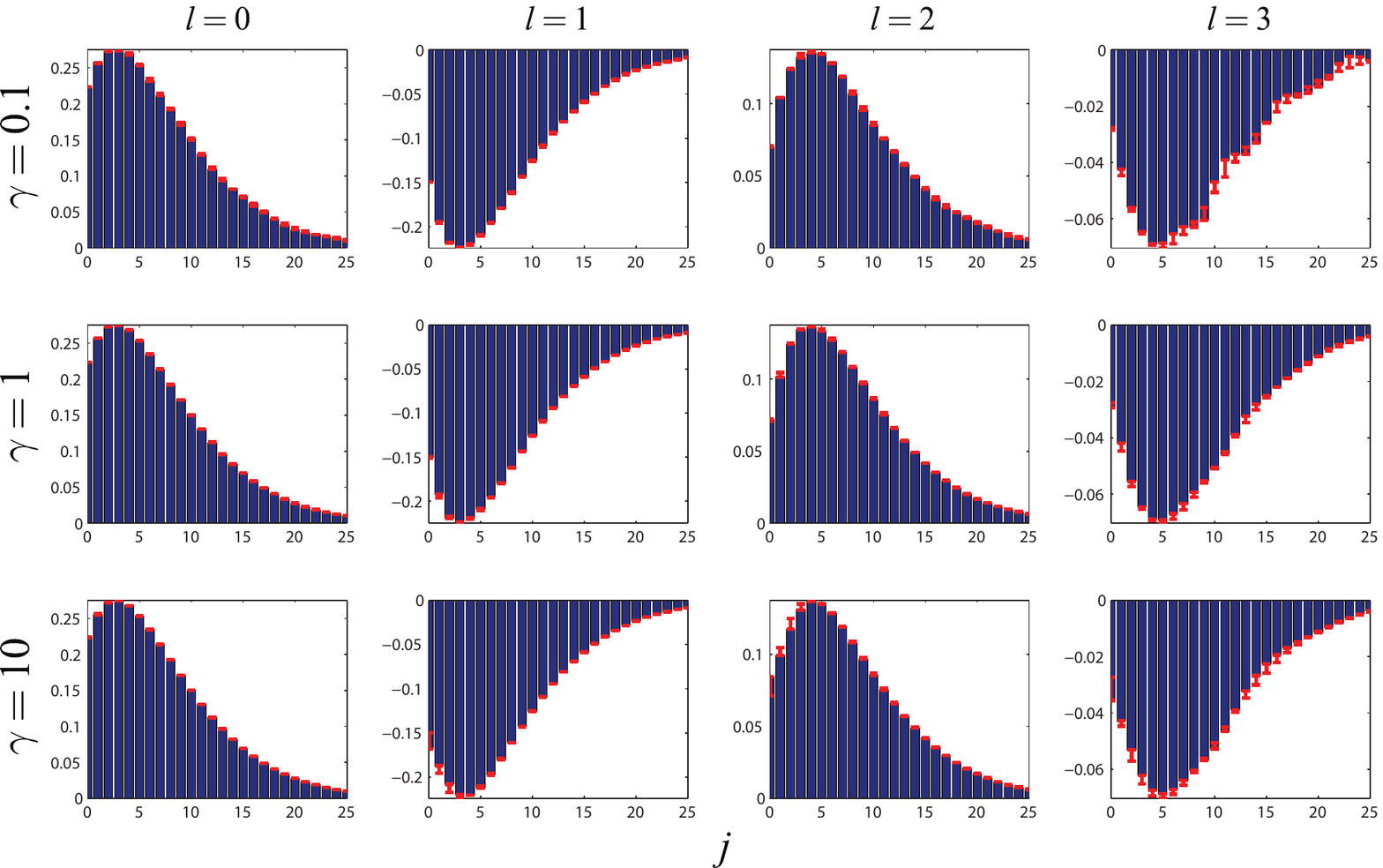}
\caption[]{Reconstructed POVM element for the no-click event of a weak-field homodyne APD (with the reflectivity of the beam-splitter of 0.5, LO intensity $|\alpha_{LO}|^2 = 5$ and quantum efficiency of the APD $60\%$) under different level of regularization $\gamma = 0.1, 1, 10$. The simulation is done for $M_p = 40$ and $f=10^5$. We demonstrate each leading diagonal separately up to $l=3$. Red bars on top of the reconstructed POVM element indicate the distance from the theoretical prediction.}
\label{fig:povm_reg}
\end{center}
\end{figure}

Most realistic optical detectors have finite efficiencies which enforces a certain degree of smoothness in their corresponding POVM representations.
If a lossy optical detector has a POVM element with non-zero amplitude $\elem{m}{n}$ it will also have a non-zero amplitude in $\elem{m+1}{n+1}, \elem{m+2}{n+2}, \dots, \elem{m+K}{n+K}$, decreasing with $K.$ If the detector has a finite efficiency $\eta$, it will impose some smoothness on the distribution $\pi_n^{j,k}$. That is because if $G(k)$ is the probability of registering $k$ photons and $H(k^{\prime})$ is the probability that $k^{\prime}$ were present, then the loss process will impose
\be
G(k)= {\displaystyle\sum\limits_{k^{\prime}}} \binom{k^{\prime}}{k}\eta^{k}(1-\eta)^{k^{\prime}-k}H(k^{\prime}).
\ee
This motivates an immediate generalization of Eq.~(\ref{eq:regcond}) to that in Eq.~(\ref{eq:sdp_rec}) as
\be
\label{eq:regcondnew}
g(\tilde{\Pi}^{(l)})=\gamma \sum_{j,n}|\pi_n^{j,j+l}-\pi_n^{j+1,j+l+1}|^2.
\ee
While $\gamma$ is a free parameter introduced into the problem for numerical smoothness, we show that the outcomes of our reconstruction procedure are fairly insensitive to the actual value of the parameter. Fig.~(\ref{fig:povm_reg}) presents the effect of the regularization condition for the reconstruction of the no-click POVM of a weak-homodyne APD with the reflectivity of the beam-splitter of 0.5, LO intensity $|\alpha_{LO}|^2 = 5$ and quantum efficiency of the APD $60\%$ (overall detection efficieny $30\%$). We vary the weight of the regularization condition for two orders of magnitude. In addition to the fidelity, we also calculate the relative error of the reconstructed POVM $||\Pi_0^{\rm rec} - \Pi_0^{\rm the}||_2/||\Pi_0^{\rm the}||_2$. The results are presented in Table~(\ref{tab:reg}), which show that the change of the reconstructed POVM elements due to the change of regularization strength is small.  This confirms that the main effect of the regularization condition is to suppress the ill-conditioning and noise while leaving the POVM fitting unaffected.

\begin{table}[ht]
\caption{\label{tab:reg}Sensitivity of the reconstruction procedure to the choice of parameter $\gamma$. No-click event of a weak-homodyne APD with the reflectivity of the beam-splitter of 0.5, LO intensity $|\alpha_{LO}|^2 = 5$ and quantum efficiency of the APD $60\%$ (overall detection efficieny $30\%$).}
\begin{tabular*}{\textwidth}{@{}l*{15}{@{\extracolsep{0pt plus 12pt}}l}}
\br
$\gamma$&Fidelity&Relative error\\
\mr
0.1&$98.38\%$&$5.54\%$\\
1&$98.32\%$&$3.33\%$\\
10&$98.36\%$&$6.88\%$\\
\br
\end{tabular*}
\end{table}

\begin{figure}[ht]
\begin{center}
\includegraphics[width=0.95\textwidth]{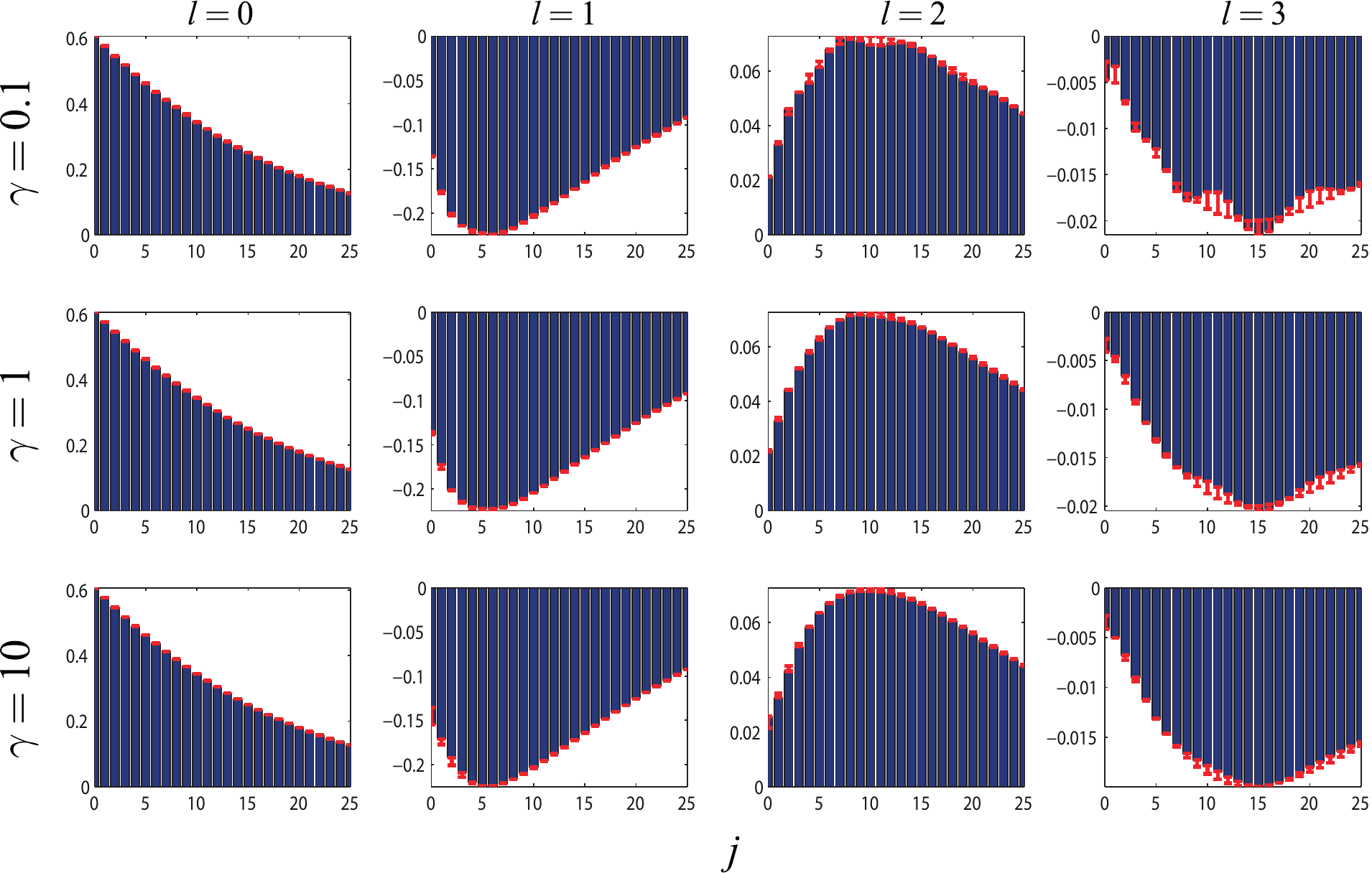}
\caption[]{Reconstructed POVM element for the no-click event of a weak-field homodyne APD (with the reflectivity of the beam-splitter of 0.5, LO intensity $|\alpha_{LO}|^2 = 5$ and quantum efficiency of the APD $20\%$) under different level of regularization $\gamma = 0.1, 1, 10$. The simulation is done for $M_p = 40$ and $f=10^5$. We demonstrate each leading diagonal separately up to $l=3$. Red bars on top of the reconstructed POVM element indicate the distance from the theoretical prediction.}
\label{fig:povm_reg_low_eff}
\end{center}
\end{figure}

\begin{figure}[ht]
\begin{center}
\includegraphics[width=0.95\textwidth]{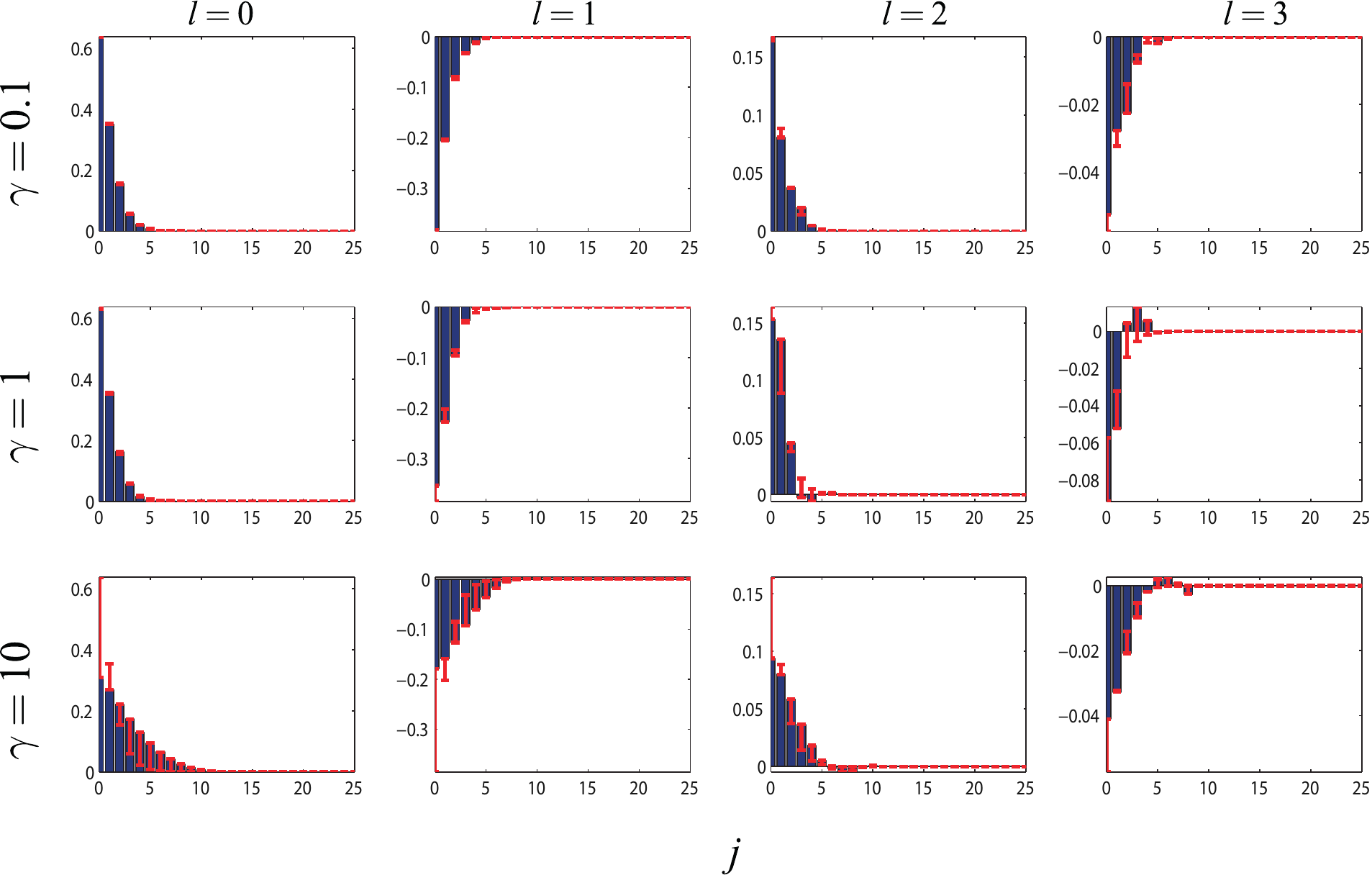}
\caption[]{Reconstructed POVM element for the no-click event of a weak-field homodyne APD (with the reflectivity of the beam-splitter of 0.1, LO intensity $|\alpha_{LO}|^2 = 5$ and quantum efficiency of the APD $90\%$) under different level of regularization $\gamma = 0.1, 1, 10$. The simulation is done for $M_p = 40$ and $f=10^5$. We demonstrate each leading diagonal separately up to $l=3$. Red bars on top of the reconstructed POVM element indicate the distance from the theoretical prediction.}
\label{fig:povm_reg_high_eff}
\end{center}
\end{figure}

As a comparison, we also calculated the reconstruction of the no-click POVM of a weak-homodyne APD with the reflectivity of the beam-splitter of 0.5, LO intensity $|\alpha_{LO}|^2 = 5$ and quantum efficiency of the APD $20\%$ (overall detection efficieny $10\%$) and that of a weak-homodyne APD with the reflectivity of the beam-splitter of 0.1, LO intensity $|\alpha_{LO}|^2 = 5$ and quantum efficiency of the APD $90\%$ (overall detection efficieny $81\%$). The results are shown in Figs.~(\ref{fig:povm_reg_low_eff}) and (\ref{fig:povm_reg_high_eff}). Calculated fidelities and relative errors are presented in Tables~(\ref{tab:reg_low_eff}) and (\ref{tab:reg_high_eff}). We can see that regularization works very well for moderate and low detection efficiencies, while its performance decreases if the detection efficiency is very high since the corresponding POVM elements are not smooth any more. On the other hand, one can infer the detection efficiency from the differences between the reconstructed results with different regularization strengths. If such difference is large, one should utilize a reduced regularization strength in the reconstruction.

\begin{table}[ht]
\caption{\label{tab:reg_low_eff}Sensitivity of the reconstruction procedure to the choice of parameter $\gamma$. No-click event of a weak-homodyne APD with the reflectivity of the beam-splitter of 0.5, LO intensity $|\alpha_{LO}|^2 = 5$ and quantum efficiency of the APD $20\%$ (overall detection efficieny $10\%$).}
\begin{tabular*}{\textwidth}{@{}l*{15}{@{\extracolsep{0pt plus 12pt}}l}}
\br
$\gamma$&Fidelity&Relative error\\
\mr
0.1&$99.85\%$&$1.34\%$\\
1&$99.87\%$&$1.28\%$\\
10&$98.87\%$&$2.32\%$\\
\br
\end{tabular*}
\end{table}

\begin{table}[ht]
\caption{\label{tab:reg_high_eff}Sensitivity of the reconstruction procedure to the choice of parameter $\gamma$. No-click event of a weak-homodyne APD with the reflectivity of the beam-splitter of 0.1, LO intensity $|\alpha_{LO}|^2 = 5$ and quantum efficiency of the APD $90\%$ (overall detection efficieny $81\%$).}
\begin{tabular*}{\textwidth}{@{}l*{15}{@{\extracolsep{0pt plus 12pt}}l}}
\br
$\gamma$&Fidelity&Relative error\\
\mr
0.1&$99.87\%$&$1.82\%$\\
1&$96.95\%$&$8.29\%$\\
10&$71.08\%$&$45.96\%$\\
\br
\end{tabular*}
\end{table}

\subsection{Reconstruction of the POVM of a weak-field homodyne APD}
\label{sec:rec_example}

\begin{figure}[ht]
\begin{center}
\includegraphics[width=0.95\textwidth]{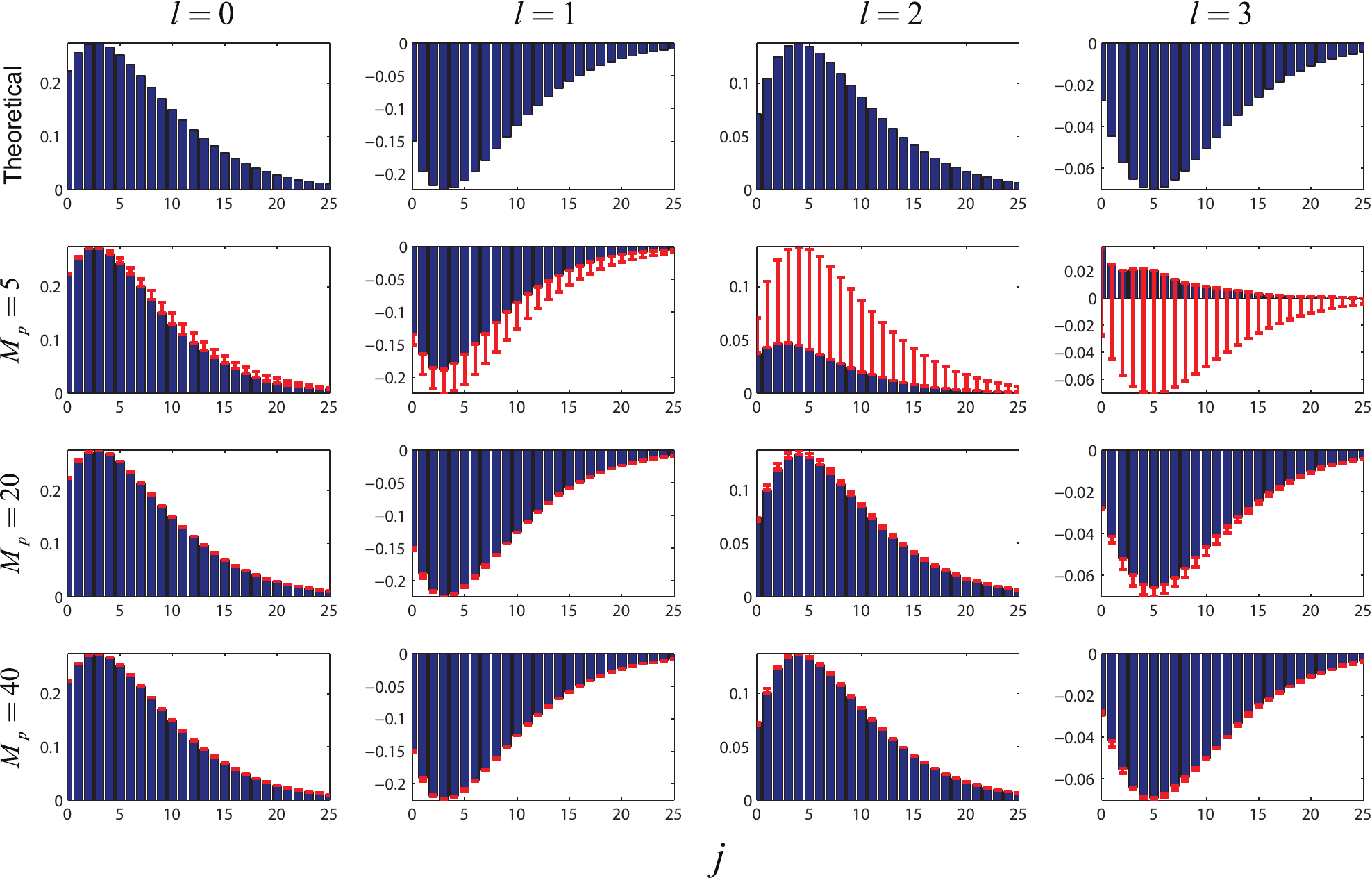}
\caption[]{Theoretical prediction and reconstructed POVM element for the no-click event of a weak-field homodyne APD. We consider three different probe phase settings $M_p = 5, 20, 40$. For each probe state we assume the experiment is run $f = 10^5$ times, and simulate the experimental fluctuation with a binomial distribution. We demonstrate each leading diagonal separately up to the $l=3$. Red bars on top of the reconstructed POVM element indicate the distance from the theoretical prediction. The results are constructed with the weight of the regularization function $\gamma = 1$.}
\label{fig:povm_phase_num}
\end{center}
\end{figure}
To discuss the performance of the recursive reconstruction method, we numerically simulate the reconstruction of the POVM set of a weak-homodyne APD with the reflectivity of the beam-splitter of 0.5, LO intensity $|\alpha_{LO}|^2 = 5$ and quantum efficiency of the APD $60\%$. We choose the intensity of the probe state $|\alpha_u|^2$ from 0 to 100 photons with a step size of 0.5 photon. This is sufficient to saturate the detector response. For each intensity we consider probe phases distributed uniformly between 0 and $2\pi$, \textit{i.e.} $\theta_{u,v} = \{0, 2\pi/M_p, \dots, 2(M_p-1)\pi/M_p\}$. Again we only consider the fluctuation induced by the random nature of the measurement process. We assume that for each probe state we run the experiment $f$ times and simulate the experimental noise by assume $f_{0|\alpha}$ is a random number with a binomial distribution. In Fig.~(\ref{fig:povm_phase_num}) we show the theoretical prediction and the reconstructed POVM for the no-click event with $M_p = 5, 20, 40$ and $f = 10^5$. To illustrate the results we show each leading diagonal separately up to $l = 3$. The reconstruction is done up to 150 photon-number component and is only displayed to 25 photon-number component for clarity. We calculate the fidelity between the reconstructed POVM $\Pi_0^{\rm rec}$ and theoretical prediction $\Pi_0^{\rm the}$
\begin{equation}
F = \left(\textrm{Tr}\left(\left(\sqrt{\Pi_0^{\rm rec}}\Pi_0^{\rm the}\sqrt{\Pi_0^{\rm rec}}\right)^{1/2} \right)\right)^2 / \textrm{Tr}\left(\Pi_0^{\rm rec}\right) \textrm{Tr}\left(\Pi_0^{\rm the}\right),
\label{eq:fidelity}
\end{equation}
which are $87.04\%$, $98.19\%$ and $98.32\%$ for $M_p = 5,$ 20 and 40 respectively. The change in fidelity can be further elucidated by the red bars on top of the reconstructed POVM element which indicate the distance from the theoretical prediction. From the results we can see that although all three phase settings give almost the same results for the principle diagonal, for higher $l$ it requires more probe phases for an accurate reconstruction. This is due to the numerical error for the calculation of the integral given in Eq.~(\ref{eq:nume_err}). This on the other hand shows a practical advantage of the recursive QDT. The probe phase setting can be decided by the elements in the POVM matrices to be reconstructed. If we are only interested in the low leading diagonals, we can greatly reduce the number of probe phases from that needed for a complete reconstruction of the POVM.

\begin{figure}[ht]
\begin{center}
\includegraphics[width=0.95\textwidth]{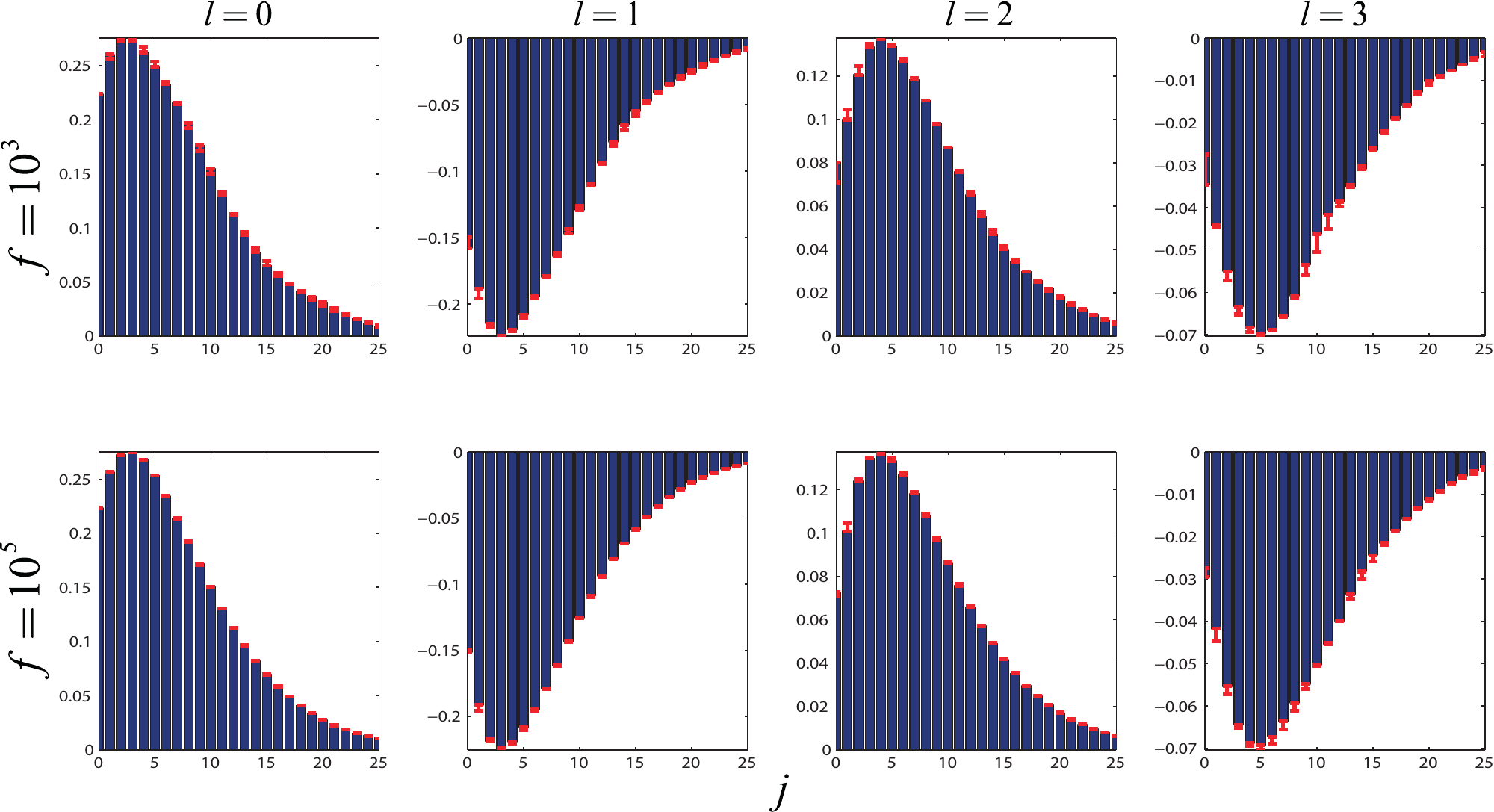}
\caption[]{Reconstructed POVM element for the no-click event of a weak-field homodyne APD under different level of experimental fluctuations. For each probe state we assume the experiment is run $f = 10^3$ and $10^5$ times, and simulate the experimental fluctuation with a binomial distribution. We demonstrate each leading diagonal separately up to the $l=3$. Red bars on top of the reconstructed POVM element indicate the distance from the theoretical prediction. The results are constructed with $M_p = 40$ and the weight of the regularization function $\gamma = 1$.}
\label{fig:povm_fluctuation}
\end{center}
\end{figure}
In Fig.~(\ref{fig:povm_fluctuation}) we show the performance of the recursive QDT under different level of experimental fluctuations. Here for each probe intensity we adjust $M_p = 40$ phases. The method discussed in Sec.~\ref{sec:pfun} requires $f=10^{10}$ to achieve a satisfactory accuracy. As a comparison, the recursive QDT is very robust against the experimental fluctuations: a decent accuracy can already be achieved for $f = 10^3$ (fidelity with the theoretical prediction $98.27\%$), with further improvement for $f = 10^5$ (fidelity $98.32\%$). Depending on the repetition rate of the detector and the laser system for LO and probe state, this requires only several millisecond to one second for each probe state.

\section{Conclusion}
\label{sec:con}
Phase-sensitive quantum-optical detectors are crucial to fully exploit the fundamental features of quantum physics and to optimally utilize optical telecommunications channels~\cite{Guha11,Kenji11,chen2011optical}. The success of these applications relies on the accurate knowledge of detectors. Yet as quantum-optical detectors become more sophisticated, normal parameters like detectivity, spectral sensitivity and noise-equivalent power are not sufficient to provide a complete specification of the detector. Moreover the complex structures of detectors and the coupling with external degrees of freedom make the conventional characterization of these detectors less feasible. Quantum detector tomography, a black-box or device-independent approach for the complete characterization of quantum detectors, provides a universal solution to this problem. Full characterization enables more flexible design and use of detectors, be they noisy, nonlinear, inefficient or operating outside their normal range. However, the large number of parameters associated with the tomography of coherent quantum detectors presents a technical challenge. This challenge is becoming increasingly typical as quantum devices grow in sophistication. In this work we present a novel recursive reconstruction algorithm to overcome this problem. Aided by numerical simulations, we have demonstrated successful reconstructions of the POVM of a weak-field homodyne APD. The results show the flexibility of the algorithm and its robustness to experimental noise. The capability to fully characterize coherent quantum-optical detectors paves the way to study genuine quantum features, including wave-particle duality, super-sensitivity \textit{etc.}, of a measurement process. It allows the benchmarking of the performance of quantum-optical detectors for various quantum applications and sheds new light on the assessment and verification of more complex detectors. We also hope that recursive quantum tomography provides an efficient procedure for quantum tomography in other quantum state and process characterization problems.

\ack
We thank G.\ Donati, T.\ J.\ Bartley, X.\ Yang, A.\ Feito, B.\ J.\ Smith, G.\ Puentes and J.\ S.\ Lundeen for assistance and fruitful discussions. This work was funded in part by EPSRC (Grant EP/H03031X/1), US EOARD (Grant 093020), EU Integrated Project Q-ESSENCE,  ERC grant TAQ, the BMBF Verbundprojekt QuoRep and the Alexander von Humboldt Foundation. I.\ A.\ W. acknowledges support from the Royal Society.

\section*{References}
\bibliographystyle{unsrt}

\end{document}